\documentclass[aps,twocolumn,preprintnumbers,nofootinbib,superscriptaddress]{revtex4}
\usepackage{amsmath} \usepackage{graphicx} \usepackage{amsfonts}
\usepackage{array} \usepackage{amsthm} \usepackage{bm}
\usepackage{palatino} \usepackage{mathpazo} 
\usepackage{supertabular}

\usepackage[breaklinks]{hyperref}
\usepackage{color}

\newcommand{\nn}{\nonumber}
\newcommand{\be}{\begin{equation}}
\newcommand{\ee}{\end{equation}}
\newcommand{\ba}{\begin{eqnarray}}
\newcommand{\ea}{\end{eqnarray}}
\newcommand{\bal}{\begin{align}}
\newcommand{\eal}{\end{align}}

\newcommand{\dd}{{\rm d}}

\newcommand{\bb}{\bibitem}

\newcommand{\al}{\alpha}

\newcommand{\bt}{\beta}
\newcommand{\ka}{\kappa}

\newcommand{\ga}{\gamma}
\newcommand{\ro}{\rho}
\newcommand{\ep}{\epsilon}
\newcommand{\si}{\sigma}
\newcommand{\ta}{\theta}

\newcommand{\Si}{\Sigma}
\newcommand{\De}{\Delta}

\newcommand{\de}{\delta}
\newcommand{\vp}{\varphi}

\newcommand{\bw}{\begin{widetext}}
\newcommand{\ew}{\end{widetext}}

\begin{document}
\title{Vacuum and nonvacuum black holes in a uniform magnetic field}

\author{Mustapha Azreg-A\"{\i}nou}
\affiliation{Ba\c{s}kent University, Engineering Faculty, Ba\u{g}l\i ca Campus, Ankara, Turkey}


\begin{abstract}
We modify and generalize the known solution for the electromagnetic field when a vacuum, stationary, axisymmetric black hole is immersed in a uniform magnetic field to the case of nonvacuum black holes (of modified gravity) and determine all linear terms of the vector potential in powers of the magnetic field and the rotation parameter.
\end{abstract}


\maketitle

\section{The magnetic field problem\label{sec1}}

A Killing vector $\xi^{\mu}$ in vacuum (no stress-energy $T_{\mu\nu}\equiv 0$) is endowed with the property of being parallel, that is proportional, to some vector potential $A^{\mu}$ that solves the source-less (no currents $J^{\mu}=F^{\mu\nu}{}_{;\nu}=(\sqrt{|g|}F^{\mu\nu}){}_{,\nu}/\sqrt{|g|}\equiv 0$) Maxwell field equations. So, $\xi^{\mu}$ is itself a solution to the same source-less Maxwell field equations. In Ref.~\cite{Wald}, this property was employed as an ansatz to determine the electromagnetic field of a \emph{vacuum}, stationary, axisymmetric, asymptotically flat black hole placed in a uniform magnetic field that is asymptotically parallel to the axis of symmetry. The ansatz stipulates that the vector potential of the solution be in the plane spanned by the timelike Killing vector $\xi^{\mu}_{t}=(1,0,0,0)$ and spacelike one $\xi^{\mu}_{\vp}=(0,0,0,1)$ of the stationary, axisymmetric black hole
\begin{equation}\label{v}
A^{\mu}=C_t(B)\xi^{\ \mu}_{t}+C_{\vp}(B)\xi^{\ \mu}_{\vp}.
\end{equation}
Since $\xi^{\ \mu}_{t}$ and $\xi^{\ \mu}_{\vp}$ are pure geometric objects, they do not encode information on the applied magnetic filed $B$; such information is encoded in the coefficients ($C_t,C_{\vp}$). Here $B$ is taken as a test field, so the metric of the stationary, axisymmetric black hole too does not encode any information on the applied magnetic field.

In this work, a spacetime metric has signature ($+,-,-,-$) and $F_{\mu\nu}=\partial_{\mu}A_{\nu}-\partial_{\nu}A_{\mu}$. For neutral and charged black holes, Eq. (4.4) of Ref.~\cite{Wald} yields
\begin{equation}\label{csn}
C_t= aB,\quad C_{\vp}=\tfrac{B}{2},
\end{equation}
and
\begin{equation}\label{csc}
C_t= aB+\tfrac{Q}{2M},\quad C_{\vp}=\tfrac{B}{2},
\end{equation}
respectively\footnote{The sign ``+" in~\eqref{csc} in front of $Q$ is due to our metric-signature choice.}, where $B$ and $Q$ are seen as perturbations, that is, if the metric of the \emph{background} black hole is that of Kerr, then $Q\xi_{t\ \mu}/(2M)$ is, up to an additive constant, the one-form $A_{\mu}\dd x^{\mu}=-Qr(\dd t-a\sin^2\ta\dd \vp)/\ro^2$ of the Kerr-Newman black hole (with $\ro^2=r^2+a^2\cos^2\ta$). It is important to emphasize this point: The potential given by~\eqref{v} and~\eqref{csc} is not an exact solution to the source-less Maxwell equations
\begin{equation}\label{VME}
F_{\al\bt;\ga}+F_{\ga\al;\bt}+F_{\bt\ga;\al}=0,\quad J^{\mu}\equiv F^{\mu\nu}{}_{;\nu}=0,
\end{equation}
if the background metric is that of the charged black hole itself. Rather, it is a solution to~\eqref{VME} if the background metric is that of the corresponding uncharged black hole.

For instance, in the Kerr background metric, the potential given by~\eqref{v} and~\eqref{csc} is a solution to~\eqref{VME}, but in the Kerr-Newman background metric the nonvanishing electric charge density $J^t$ and the $\vp$ current density $J^{\vp}$ expand in powers of $Q$ as
\begin{align*}
&J^t=\frac{(\sqrt{|g|}F^{t\nu}){}_{,\nu}}{\sqrt{|g|}}=
\frac{2aBQ^2(a^2+r^2\cos^2\ta)}{(r^2+a^2\cos^2\ta)^3}+\mathcal{O}(Q^3),\\
&J^{\vp}=\frac{(\sqrt{|g|}F^{\vp\nu}){}_{,\nu}}{\sqrt{|g|}}=
\frac{BQ^2[a^2(2+\cos^2\ta)-r^2]}{(r^2+a^2\cos^2\ta)^3}+\mathcal{O}(Q^3),
\end{align*}
which are zero to first order only. Even if rotation is suppressed ($a=0$), $J^t$ is still nonzero:
\[J^t=\frac{-2Q^5}{(2Mr-Q^2)^3r},\] and its integral charge is also nonzero. Where does this electric charge density come from (the only existing electric charge is that of the black hole, which is confined inside the event horizon)? Because of the conservation of the total electric charge, the application of a uniform magnetic field does not generate current densities $J^{\mu}$ outside the event horizon.

Thus, as far as $Q$ is considered as a perturbation, the potential given by~\eqref{v} and~\eqref{csc} remains a good approximation for many astrophysical purposes. However, this fails to be the case if one is interested in the accretion phenomena that take place in the vicinity of the innermost stable circular orbit (ISCO) whose radius approaches that of the event horizon, for there the currents ($J^t,J^{\vp}$) cannot be neglected. One of the purposes of this paper is to provide an ``exact" formula for the vector potential of a vacuum \emph{charged} black hole immersed in a uniform magnetic filed parallel to its axis of symmetry. The purpose extends to include nonvacuum charged and uncharged black holes.

\section{The solution\label{sec2}}

\subsection{General considerations}

The first thing we want to show in this section is that the expressions~\eqref{csn} are universal leading terms of more elaborate formulas for ($C_t,C_{\vp}$). The determination of these leading terms is purely geometrical and only depends on the asymptotic behavior of the metric of a (vacuum or nonvacuum), stationary, axisymmetric, \emph{asymptotically flat} black hole. Here by asymptotical flatness we mean that, at spatial infinity, $T_{\mu\nu}\to 0$ ensuring Ricci-flatness $R_{\mu\nu}\to 0$ and hence $F^{\mu\nu}{}_{;\nu}\to 0$. The metric of such a black hole approaches that of a rotating star as the radial coordinate tends to infinity
\begin{align}\label{gas}
\dd s^2 \simeq & \Big(1-\frac{2M}{r}\Big)\dd t^2-\frac{1}{1-2M/r}\,\dd r^2
+\frac{4aM \sin ^2\theta}{r}\,\dd t\dd \vp\nn\\
\quad & -r^2(\dd \ta^2+\sin ^2\theta\,\dd \vp^2).
\end{align}
Using this, a direct integration of~\eqref{VME} yields the leading terms
\begin{equation}\label{Fas}
F_{r\vp}\simeq -2C_{\vp}r\sin^2\ta,\quad F_{\ta\vp}\simeq -2C_{\vp}r^2\sin\ta\cos\ta,
\end{equation}
which were derived previously for the Kerr black hole (Eq.~(6) of Ref.~\cite{Aliev}). This behavior is universal and applies to all (vacuum or nonvacuum), stationary, axisymmetric, asymptotically flat black holes. The leading terms proportional to $C_t$ vanish as $1/r$. Eqs.~\eqref{Fas} are the expressions of the electromagnetic tensor $F_{\mu\nu}$, expressed in spherical coordinates, of a uniform magnetic field, of strength $B$, parallel to the $z$ axis provided we take
\begin{equation}\label{cvp}
C_{\vp}=\tfrac{B}{2}.
\end{equation}

There are three points to emphasize in the above derivation. First, notice that the derivation of~\eqref{cvp} is valid whether the black hole is neutral or charged, for the presence of a charge, would certainly modify the expansion in~\eqref{gas}, but would not modify the leading terms in~\eqref{Fas}. Second, we have made no assumption on the nature of the electrodynamics (linear or nonlinear) describing $F_{\mu\nu}$. Thus, the value of $C_{\vp}=B/2$ applies equally to black holes with a linear electromagnetic source as well as to, generally regular (singularity-free), charged black holes with a nonlinear electromagnetic source provided they are asymptotically flat with a vanishing stress-energy at spatial infinity yielding $R_{\mu\nu}\to 0$ which, in turn, yields $J^{\mu}\to 0$. Thirdly, we have made no link to general relativity nor to any of its modifications and extensions. Hence, no matter the theory of relativity describing the geometry and physics of spacetime and matter, the value of $C_{\vp}=B/2$ applies to all stationary, axisymmetric, asymptotically flat solutions (black holes, wormholes, etc ...) if they are placed in a uniform magnetic field $B$ parallel to the symmetry axis. An instance of application of these arguments is the case of a non-Kerr black hole where it was shown that~\eqref{cvp} applies~\cite{B1}.

As for $C_t$, since we are dealing with asymptotically flat solutions, we are implicitly assuming that at spatial infinity linear electrodynamics~\eqref{VME} is sufficient for the description of the electromagnetic field. The charge of the black hole is given by the surface integral
\begin{equation}\label{Q1}
Q=\frac{1}{8\pi}\int_{\partial\Si} F^{\mu\nu}\dd \Si_{\mu\nu},
\end{equation}
and this does not depend on the surface $\partial\Si$. This is the initial charge, if any, of the black hole. If $\partial\Si$ is a surface of fixed radial coordinate $r$, then $\Si_{\mu\nu}=e_{\mu\nu\ta\vp}\sqrt{|g|}\dd\ta\dd\vp$ (where the totally antisymmetric symbol $e_{\mu\nu\al\bt}$ is such that $e_{tr\ta\vp}=+1$). In the limit $r\to\infty$, the metric of the black hole~\eqref{gas} is known yielding
\begin{equation}\label{Q2}
    Q=2(C_t-aB)M+\mathcal{O}(1/r),
\end{equation}
where we have used~\eqref{cvp}. In the limit $r\to\infty$, we rederive the first equation in~\eqref{csn} [resp. in~\eqref{csc}] if the black hole is not charged (resp. charged). Corrections to~\eqref{gas} do not affect the leading term in~\eqref{Q2}.

\subsection{Generalizing the ansatz}

In more general situations where the Killing vector $\xi^{\mu}$ is not a solution to the source-less Maxwell field equations, a linear combinations of all Killing vector with constant coefficient, which is also a Killing vector as in~\eqref{v}, may fail to be a solution to the source-less Maxwell field equations~\eqref{VME}. If the coefficients were taken as functions of the coordinates, it would be possible to determine them upon solving~\eqref{VME}. This would allow to generalize Wald's formulas~(\ref{csn}, \ref{csc}) and this is the purpose of this work. These generalizations are useful for a consistent analysis of (un)charged-particle dynamics around black holes.

It is worth mentioning that some specific generalizations of Wald's formulas~(\ref{csn}, \ref{csc}) to metrics not obeying the expansion~\eqref{gas} have been made but no general formulas were derived. For instance, it was shown~\cite{B2} that~\eqref{csn} remains valid for the Kerr-Taub-NUT black hole but neither~\eqref{csn} nor~\eqref{csc} remains valid for black holes in Ho\v{r}ava-Lifshitz gravity and in braneworld where extensions to Wald's formulas~(\ref{csn}, \ref{csc}) have been performed in~\cite{B3} and~\cite{B4}, respectively.

For the remaining part of this section, we rewrite the ansatz~\eqref{v} in terms of ($c_t,c_{\vp}$) where $C_t=aB+c_t$ and $C_{\vp}=(B/2)+c_{\vp}$
\begin{equation}\label{v2}
\hspace{-1mm}A^{\mu}=[aB+c_t(r,\ta,a,B)]\xi^{\ \mu}_{t}+[\tfrac{B}{2}+c_{\vp}(r,\ta,a,B)]\xi^{\ \mu}_{\vp},
\end{equation}
where the dependence on $\ta$ is not related to that on the rotation parameter $a$; that is, if rotation is suppressed ($a=0$), the coefficients ($c_t,c_{\vp}$) may still depend on $\ta$.

Now, we need a metric formula to complete the task of integrating the source-less equations~\eqref{VME}. There is no generic metric that describes all rotating black holes (and wormholes) nor a generic metric for static solutions. Nonetheless, the G\"urses-G\"ursey metric~\cite{GG}
\begin{align}\label{m1}
\dd s^2 =&\Big(1-\frac{2f}{\ro^2}\Big)\dd t^2-\frac{\ro^2}{\De}\,\dd r^2
+\frac{4af \sin ^2\theta}{\ro^2}\,\dd t\dd \vp\nn\\
\quad & -\ro^2\dd \ta^2-\frac{\Si\sin ^2\theta}{\ro^2}\,\dd \vp^2,
\end{align}
where
\begin{align}\label{m2}
&\ro^2=r^2+a^2\cos^2\ta,\quad \De(r) = r^2-2f+a^2, \nn\\
&\Si =(r^2+a^2)^2-a^2\De\sin^2\ta,
\end{align}
describes a variety of solutions including (A) Schwarzschild ($a=0$), Reissner-Nordstr\"om ($a=0$), Kerr, Kerr-Newman metrics, the Schwarzschild-MOG ($a=0$) and Kerr-MOG black holes of the modified gravity (MOG)~\cite{MOG,MOG2}, and their trivial generalizations the Reissner-Nordstr\"om-MOG ($a=0$) and Kerr-Newman-MOG, and some phantom Einstein-Maxwell-dilaton black holes~\cite{phantom,multi}. It also includes (B) nonrotating regular black holes~\cite{regular1}-\cite{azth} and their rotating counterparts~\cite{gen1,gen2,gen3} as well as some nonrotating black holes of f(R) and f(T) gravities~\cite{m3,T}.

An \emph{important property} of the metric~\eqref{m1}, \eqref{m2} is the following. In the most generic case, where $f$ and ($c_t,c_{\vp}$)~\eqref{v2} are still arbitrary functions and for all values of the rotation parameter $a$ , the first set of source-less equations~\eqref{VME} and two equations of the second set are satisfied:
\begin{equation}\label{pp1}
F_{\al\bt;\ga}+F_{\ga\al;\bt}+F_{\bt\ga;\al}\equiv 0,\quad F^{r\nu}{}_{;\nu}\equiv 0,\quad F^{\ta\nu}{}_{;\nu}\equiv 0,
\end{equation}
leaving only two differential equations to solve:
\begin{equation}\label{pp2}
F^{t\nu}{}_{;\nu}=0,\quad F^{\vp\nu}{}_{;\nu}=0.
\end{equation}
Moreover, if rotation is suppressed ($a=0$) then the coefficient ($c_t,c_{\vp}$) no longer depend on $\ta$.
\begin{table}[h]
\caption{{\footnotesize The constants ($f_1,f_2$) defining the function $f(r)$~\eqref{f} for the set (A) of singular nonrotating and rotating (vacuum and nonvacuum) black holes in terms of the mass $M$, the electric charge $Q$, and the universal ratio $\ka$ of the scalar charge $Q_s$ to the mass (of any particle and, particularly, of the MOG black hole). The Newtonian gravitational constant $G_N=1$. The black hole named ``Reissner-Nordstr\"om-MOG" and ``Kerr-Newman-MOG" were not derived in Ref.~\cite{MOG} since the author was not interested in astrophysical electrically charged solutions; these are trivial generalizations of the Schwarzschild-MOG and Kerr-MOG derived in Eqs.~(11) and~(35) of Ref.~\cite{MOG}, respectively. \textsc{Nomenclature:} ``V" for ``vacuum solution" (Ricci-flat: $R_{\mu\nu}=0$), ``NV" for ``nonvacuum solution" (non-Ricci-flat: $R_{\mu\nu}\neq 0$), ``N" for ``neutral" (electrically uncharged), and ``C" for ``electrically charged".}}\label{Tab1}
{\footnotesize
\begin{tabular}{|l|c|c|c|}
  \hline
  &  &  & \vspace{-0.3cm} \\
  {Black hole} & $f_1$ & $f_2$ & {State} \\
  &  &  & \vspace{-0.3cm} \\
  \hline
  \hline
  Schwarzschild & $M$ & $0$ & V, N \\
  \hline
  Reissner-Nordstr\"om & $M$ & $-Q^2/2$ & NV, C  \\
  \hline
  Kerr & $M$ & $0$ & V, N  \\
  \hline
  Kerr-Newman & $M$ & $-Q^2/2$ & NV, C  \\
  \hline
  Schwarzschild-MOG & $(1+\ka^2)M$ & $-(1+\ka^2)\ka^2M^2/2$ & NV, N  \\
  \hline
  Reissner-Nordstr\"om- &  &  &  \\
  MOG & $(1+\ka^2)M$ & $-(1+\ka^2)(\ka^2M^2+Q^2)/2$ & NV, C  \\
  \hline
  Kerr-MOG & $(1+\ka^2)M$ & $-(1+\ka^2)\ka^2M^2/2$ & NV, N  \\
  \hline
  Kerr-Newman-MOG & $(1+\ka^2)M$ & $-(1+\ka^2)(\ka^2M^2+Q^2)/2$ & NV, C  \\
  \hline
  Phantom  &  &  &  \\
  Reissner-Nordstr\"om & $M$ & $Q^2/2$ & NV, C \\
  \hline
\end{tabular}}
\end{table}

For the set (A) of singular solutions, the function $f(r)$ is linear of the form
\begin{equation}\label{f}
    f(r)=f_1r+f_2,
\end{equation}
where the constants ($f_1,f_2$) do not depend on the rotation parameter $a$. Table~\ref{Tab1} gives the values of these constants for the set (A) of black hole solutions.

The electromagnetic field of regular black holes does not satisfy the linear source-less Maxwell equations~\eqref{VME}, except asymptotically, and their function $f$ admits the expansion: $f_1r+f_2+\sum_{i=1} c_i/r^i$. In this work we will not deal with metrics of regular black holes.

\section{The nonrotating case: $\mathbf{a=0}$}
We will first deal with the nonrotating case, where the black hole has the spherical symmetry, setting $a=0$ the ansatz~\eqref{v2} reduces to:
\begin{equation}\label{v3}
A^{\mu}=c_t(r,B)\xi^{\ \mu}_{t}+[\tfrac{B}{2}+c_{\vp}(r,B)]\xi^{\ \mu}_{\vp},
\end{equation}
for ($c_t,c_{\vp}$) do not depend on $\ta$. With this last form of the ansatz, $F^{t\nu}{}_{;\nu}=0$ implies
\begin{multline}\label{t1}
r^2 (r^2-2 f) c_t''+2 r (r^2+2 f-2 r f') c_t'\\-2 [2 f-r (2 f'-r f'')] c_t=0,
\end{multline}
where the prime denotes derivative with respect to $r$. This is solved by
\begin{equation}\label{t2}
c_t=\frac{K_1r}{r^2-2f}+\frac{K_2r^2}{r^2-2f}=\frac{K_1}{rg_{tt}}+\frac{K_2}{g_{tt}},
\end{equation}
where ($K_1,K_2$) are integration constants. Independently of the value of $K_2$, the integral~\eqref{Q1} performed on a sphere of radius $r$ yields $K_1=-Q$ if the black hole is charged or $K_1=0$ if the black hole is neutral. We can take $K_2=0$, for this is an additive constant in the expression of the one-form $A_{\mu}\dd x^{\mu}=-(Q/r)\dd t+K_2\dd t +$ terms proportional to $\dd\vp$. The result~\eqref{t2} was expected and constitutes a correction to Wald's term $Q/(2M)$ in~\eqref{csc}.

Now, $F^{\vp\nu}{}_{;\nu}=0$ reduces to the two equivalent differential equations
\begin{align}
&(r^4-2 r^2 f) c_{\varphi }''+(4 r^3-4 r f-2 r^2 f') c_{\varphi }'+4 (f-r f') c_{\varphi }\nn\\
&=2 B (r f'-f),\\
\label{p1}&4(f-r f') C_{\vp}+[(r^4-2 r^2 f) C_{\vp}']'=0,
\end{align}
where $C_{\vp}=(B/2)+c_{\vp}$, as defined earlier~\eqref{v2}. In the generic case, where $f$ is any function of $r$, a closed-form solution to~\eqref{p1} does not exist. For $f$ linear~\eqref{f}, we obtain the solution
\begin{align}
\label{p2}&C_{\vp}=\frac{L_1 (r^2+2 f_2)}{2 r^2}\\
&+\frac{L_2 (r^2+2 f_2) \Big[\frac{\sqrt{f_1^2+2 f_2}(r+f_1)}{r^2+2 f_2}-\tfrac{1}{2}\ln \Big(\tfrac{r+\sqrt{f_1^2+2 f_2}-f_1}{r-\sqrt{f_1^2+2
f_2}-f_1}\Big)\Big]}{8 (f_1^2+2 f_2)^{3/2} r^2},\nn
\end{align}
where ($L_1,L_2$) are integration constants. We immediately get rid of $L_2$, for the physical electromagnetic tensor $F_{\mu\nu}$ would diverge at the horizons $r_{\pm}=f_1\pm \sqrt{f_1^2+2 f_2}$. At spatial infinity, $C_{\vp}$ reduces to $B/2$~\eqref{cvp} yielding $L_1=B$ and
\begin{equation}\label{p3}
C_{\vp}=\frac{B}{2}+\frac{B f_2}{r^2}.
\end{equation}

For the linear case~\eqref{f}, we have obtained the solution for the vector potential
\begin{equation}\label{solg}
A^{\mu}=-\frac{Q}{rg_{tt}}\xi^{\ \mu}_{t}+\frac{B}{2}\Big(1+\frac{2 f_2}{r^2}\Big)\xi^{\ \mu}_{\vp}.
\end{equation}
For the Schwarzschild, normal or phantom Reissner-Nordstr\"om, Schwarzschild-MOG, and Reissner-Nordstr\"om-MOG
black holes we obtain respectively
\begin{align}\label{solapp}
&A^{\mu}=\tfrac{B}{2}\xi^{\ \mu}_{\vp},\\
&A^{\mu}=-\tfrac{Q}{rg_{tt}}\xi^{\ \mu}_{t}+\tfrac{B}{2}\Big(1-\tfrac{Q^2}{r^2}\Big)\xi^{\ \mu}_{\vp},\\
&A^{\mu}=\tfrac{B}{2}\Big(1-\tfrac{(\ka^2+1)\ka^2M^2}{r^2}\Big)\xi^{\ \mu}_{\vp},\\
&A^{\mu}=-\tfrac{Q}{rg_{tt}}\xi^{\ \mu}_{t}+\tfrac{B}{2}\Big(1-\tfrac{(\ka^2+1)(\ka^2M^2+Q^2)}{r^2}\Big)\xi^{\ \mu}_{\vp}.
\end{align}
Notice that it is the correction inside the parentheses, with respect to Wald's formula~\ref{csc}, that ensures the satisfaction of the source-less Maxwell equations~\eqref{VME}. According to MOG theory~\cite{MOG,MOG2}, $\ka$ is the ratio of the scalar charge $Q_s$ of any particle to its mass $m$: $\ka\equiv Q_s/m$. This ratio is postulated to be universal and it is the same for all particles and massive bodies as black holes.

There is a couple of facts and conclusions to draw from~\eqref{solg}.
\begin{enumerate}
  \item We assume that the applied magnetic field is directed in the positive $z$ axis ($B>0$) and consider an uncharged black hole. On a charged particle, of electric charge $q$, the applied Lorentz magnetic force in the $\vp$ direction takes the form
      \begin{equation}\label{force}
      q F_{\si}{}^{\vp}\tfrac{\dd x^{\si}}{\dd s}=\tfrac{qB}{r^2}~\tfrac{\dd r}{\dd s}+qB\cot\ta~\tfrac{\dd \ta}{\dd s}+\tfrac{2qBf_2\cot\ta}{r^2}~\tfrac{\dd \ta}{\dd s}.
      \end{equation}
      The third term is an extra term that would be missing had we applied Wald's formula~\eqref{csn}. Recall that Wald's formulas~\eqref{csn} and~\eqref{csc} apply only to Schwarzschild and Kerr black holes. Notice that this extra force exists for neutral black holes of the MOG theory where $f_2\neq 0$ (Table~\ref{Tab1}). Since $f_2<0$, the extra force drives positively charged particles to accelerate (respectively, negatively charged particles to decelerate) in the increasing $\vp$ direction if they are approaching the axis of symmetry ($\dd \ta/\dd s<0$). This effect is new and cannot be neglected in the vicinity of the event horizon or the ISCO.

      There is a similar extra magnetic term
      \begin{equation}
        -\tfrac{2qBf_2\sin\ta\cos\ta}{r^2}~\tfrac{\dd \vp}{\dd s},
      \end{equation}
      in the applied Lorentz magnetic force in the $\ta$ direction.

      These extra forces are attributable to the minimum coupling of the magnetic field, via the covariant derivative~\eqref{VME}, to the stress-energy.
  \item In fact, this effect, which was masked by the approximation made in~\eqref{csc}, exists for all nonvacuum charged black holes no matter the nature of the stress-energy is. In our restriction~\eqref{m1}, this effect has been derived in the linear case~\eqref{f} but it should apply to neutral or charged generic configurations too where $f$ expands as $f_1r+f_2+\text{power series in }1/r$, as are the cases with the Ay\'on-Beato--Garc\'{\i}a static black hole~\cite{reg}, the solutions derived in~\cite{azth,non}, and the black holes of the $f(T)$~\cite{T} and $f(R)$~\cite{R} gravities.
  \item In the derivation of both formulas~\eqref{csc} and~\eqref{solg} it was assumed that the magnetic field is a test field, thus neglecting its backreaction. The Einstein field equations are only approximately satisfied. If $T^{\mu}_{\ \nu\ (0)}$ is the stress-energy corresponding to $B=0$, then, for instance, in the nonrotating case all the new electromagnetic extra terms\footnote{\[T^{\mu}_{\ \nu\ \text{EM}}=-\tfrac{1}{4\pi}\big(F^{\mu\al}F_{\nu\al}-\tfrac{1}{4}\de^{\mu}_{\ \nu}F^{\al\bt}F_{\al\bt}\big).\]}, $T^{\mu}_{\ \nu\ \text{EM}}$, added to $T^{\mu}_{\ \nu\ (0)}$, when $B$ is applied, are proportional to $B^2$, if the background black hole is uncharged. The linear approximation~\eqref{solg} is valid if the end-behavior as $r\to\infty$ of $T^{\mu}_{\ \nu\ \text{EM}}$ is much smaller than that of $T^{\mu}_{\ \nu\ (0)}$ yielding the constraints
      \begin{equation}\label{b1}
        B^2|f_2|r^2\ll |f_2|,\quad B^2f_1r^3\ll |f_2|.
      \end{equation}
      While the intergalactic magnetic field is supposed to be weak but these constraints show that~\eqref{solg} fails to provide a valid approximation as $r\to\infty$. This very conclusion was stated in Ref.~\cite{Ernst1} concerning Wald's linear approximation~\eqref{csc}: ``... Wald's solution must break down as $r\to\infty$, since the linearized solution is asymptotically flat, ...". This conclusion extends also to charged solutions where, besides the constraint $T^{t}_{\ \vp\ \text{EM}}\propto BQ/r\ll 1$, we have similar constraints to~\eqref{b1}
      \begin{equation}\label{b2}
        B^2|f_2|r^2\ll |f_2|+\tfrac{Q^2}{2},\quad B^2f_1r^3\ll |f_2|+\tfrac{Q^2}{2}.
      \end{equation}
  \item Ernst devised a procedure for generating Melvin-type magnetic universes from Einstein-Maxwell solutions~\cite{Ernst2}. Being non-flat with a Melvin~\cite{Melvin} asymptotic behavior, these solutions are useless for many astrophysical applications except in regions with strong magnetic field. The substitutes to Ernst's universes are the asymptotically flat solutions immersed in weak magnetic fields with linear vector potentials~\eqref{csc} and~\eqref{solg}. All linear terms in powers of $B$ have been determined in~\eqref{solg}, however, we do not expect the next quadratic corrections to have simple mathematical structures even in the nonrotating case.
\end{enumerate}

\section{The rotating case: $\mathbf{a\neq 0}$}
This case is more involved. First of all, note that rotation mixes the electric field of the background black hole with the test magnetic field~\cite{JP}. In this case it would not be possible to set the conditions constraining the test magnetic field, as we did in~\eqref{b1} and~\eqref{b2}, unless the effects of rotation are weak, in which case the two constraints~\eqref{b1} and~\eqref{b2} remain valid to the linear approximation in the rotation parameter $a$. On the other hand, if the sources of the test magnetic field carry an electric charge density $\ro$, the dragging effects, which are proportional to $a$, cause the sources to accelerate in the geometry of the background black hole and thus enhance the magnetic field they generate\footnote{It is well known that a charged rotating disk generates at its center a magnetic field linearly proportional to the disk's angular velocity. This extends to the case of an electromagnetic field around a Kerr black hole where the magnetic field, Eq.~(29) of Ref.~\cite{JP}, is proportional to $aq$ with $q$ being the total charge of the electromagnetic source.}. Since in this work we are performing a general analysis in which we ignore the sources of the test magnetic field, it is safe to perform the analysis to the linear approximation in $a$ to ensure that the generated $B$ remains small, and the analysis remains valid, in all cases.

We seek a linear solution in $a$ of the form~\eqref{v2}
\begin{multline}\label{rc1}
A^{\mu}=\Big[\frac{-Qr}{r^2-2f}+a[B+T(r,\ta,B)]\Big]\xi^{\ \mu}_{t}\\
+\Big[\frac{B}{2}\Big(1+\frac{2f_2}{r^2}\Big)+a \Phi(r,\ta,B)\Big]\xi^{\ \mu}_{\vp},
\end{multline}
where we have written ($c_t,c_{\vp}$) and as sums of the new nonrotating contributions~\eqref{solg} and linear terms in $a$. It is understood from~\eqref{rc1} that we restrict ourselves to the case $f(r)=f_1r+f_2$~\eqref{f}.

The equations to solve~\eqref{pp2} take the following forms in the $a$-linear approximation
\begin{align}
&F^{t\nu}{}_{;\nu}\propto [12 B f_2 (f_1 r+2 f_2)-4 B f_2 (r^2+3 f_1 r+6 f_2) \cos^2\ta\nn\\
&+r^4 (r^2-2 f) T^{(2,0)}+r^4\sin^2\ta T^{(0,2)}+2 r^3 (r^2+2 f_2) T^{(1,0)}\nn\\
\label{rc2}&-2 r^4 \cos\ta T^{(0,1)}-4 f_2 r^2T]a+\mathcal{O}(a^2),\\
&F^{\vp\nu}{}_{;\nu}\propto [\tfrac{2 Q (f_1 r^3+6 f_2 r^2-6 f_1 f_2 r-4 f_2^2)}{r^3 (r^2-2 f)^2}\nn\\
&+(r^2-2 f) \Phi^{(2,0)}+\sin^2\ta \Phi^{(0,2)}+(4 r-6 f_1-\tfrac{4 f_2}{r}) \Phi^{(1,0)}\nn\\
\label{rc3}&-4\cos\ta \Phi^{(0,1)}+\tfrac{4 f_2 \Phi}{r^2}]a+\mathcal{O}(a^2),
\end{align}
where, for instance, $T^{(m,n)}\equiv \frac{\partial^{m+n}T}{\partial r^m\partial x^n}$ with $x\equiv \cos\ta$. Looking for solutions of the form
\begin{equation*}
T=T_1(r)P_0(x)+T_2(r)P_2(x)\quad\text{and}\quad \Phi\equiv \Phi_1(r)P_0(x),
\end{equation*}
where $P_0(x)=1$ and $P_2(x)=-\frac{1}{2}+\frac{3x^2}{2}$ are the Legendre polynomials, we were lead to the particular solutions\footnote{By particular solutions we mean we have set the integration constants to their specific values so that the r.h.s of~\eqref{Q1} reduces to $Q+O(a^2)$. For instance, the expression
\begin{equation*}
T(r,\ta) = \frac{Bf_2\sin^2\ta}{r^2}+\frac{c r}{r^2-2f},
\end{equation*}
which a solution to~\eqref{rc2}, would yield $Q-c a+O(a^2)$, so we had to choose $c =0$.}
\begin{equation}\label{rc4}
T(r,\ta) = \frac{Bf_2\sin^2\ta}{r^2},\quad \Phi(r) = -\frac{Q}{r(r^2-2f)}.
\end{equation}

Finally, the expression of $A^{\mu}$ takes the form
\begin{multline}\label{rc5}
A^{\mu}=\Big[\frac{-Qr}{r^2-2f}+aB\Big(1+\frac{f_2\sin^2\ta}{r^2}\Big)\Big]\xi^{\ \mu}_{t}\\
+\Big[\frac{B}{2}\Big(1+\frac{2f_2}{r^2}\Big)-\frac{Qa}{r(r^2-2f)}\Big]\xi^{\ \mu}_{\vp}.
\end{multline}
To the linear approximation in $a$, the nonvanishing components of the electromagnetic tensor remain finite on the horizons
\begin{align}\label{fld1}
&F_{tr}=-\tfrac{Q}{r^2}-\tfrac{B a [4 f_2+3 f_1 r+(4 f_2+f_1 r) \cos  2 \theta ]}{2 r^3},\nn\\
&F_{t\theta}=-\tfrac{B a (2 f_2+f_1 r) \sin  2 \theta }{r^2},\nn\\
&F_{r\vp}=-B r \sin ^2 \theta -\tfrac{Q a \sin ^2 \theta }{r^2},\\
&F_{\theta \varphi }=-\tfrac{B }{2}(2 f_2+r^2) \sin  2 \theta +\tfrac{Q a \sin  2 \theta }{r},\nn
\end{align}
and
\begin{align}
&F^{t r}=\tfrac{Q}{r^2}+\tfrac{B a (2 f_2+f_1 r) (1+3 \cos  2 \theta )}{2 r^3},\nn\\
&F^{t \theta }=\tfrac{B a f_2 \sin  2 \theta }{r^4},\nn\\
\label{rc6}&F^{r \varphi }=-\tfrac{B (r^2-2 f)}{r^3}-\tfrac{Q a}{r^4},\\
&F^{\theta \varphi }=-\tfrac{B (2 f_2+r^2) \cot \theta }{r^4}+\tfrac{2 Q a \cot \theta }{r^5}.\nn
\end{align}
Using these expressions, it is straightforward to check that the r.h.s of~\eqref{Q1} reduces to $Q+O(a^2)$. The two invariants of the electromagnetic field,
\begin{align}
&F_{\mu\nu}F^{\mu\nu}=-\tfrac{Q^2}{r^4}+\tfrac{B^2[r^4-f_1r^3+f_2r^2+2f_2^2+(f_1r^3+3f_2r^2+2f_2^2)\cos2\ta]}{r^4}\nn\\
&\qquad\qquad -\tfrac{QBa[r^2+3f_1r+8f_2+(3r^2+f_1r+8f_2)\cos2\ta]}{r^5},\\
&\textsuperscript{*}F_{\mu\nu}F^{\mu\nu}=-\tfrac{4BQ(r^2+2f_2)\cos\ta}{r^4}+\tfrac{8Q^2a\cos\ta}{r^5}\\
&-\tfrac{2 B^2 a [f_1 r^3+6 f_1 f_2 r+8 f_2^2+(3 f_1 r^3+8 f_2 r^2+2 f_1 f_2 r+8 f_2^2) \cos  2 \theta ] \cos  \theta  }{r^5}\nn,
\end{align}
(where \textsuperscript{*}$F^{\mu\nu}=\tfrac{1}{2}\ep^{\mu\nu\al\bt}F_{\al\bt}$ and $\ep_{\mu\nu\al\bt}=e_{\mu\nu\al\bt}\sqrt{|g|}$ is the the totally antisymmetric tensor) too remain finite to the linear approximation in $a$.

As explained in the introduction section~\ref{sec1}, Wald's formulas~\eqref{csn} and~\eqref{csc} apply only to Schwarzschild and Kerr black holes where $f(r)=f_1r=Mr$. However, had we tried to apply~\eqref{csc} to charged black holes with $f(r)=Mr+f_2$~\eqref{f} and metric of the form~(\ref{m1},\ref{m2}), we would obtain
\begin{align}
&F_{_{\text{W}}}^{t r}=\tfrac{Q}{r^2}+\tfrac{2 f_2Q}{M r^3}+\tfrac{B a [4 f_2+M r+(4 f_2+3 f_1 r) \cos  2 \theta ]}{2 r^3},\nn\\
&F_{_{\text{W}}}^{t \theta }=0,\nn\\
\label{rc7}&F_{_{\text{W}}}^{r \varphi }=-\tfrac{B (r^2-2 f)}{r^3}-\tfrac{Q a (2 f_2+M r)}{M r^5},\\
&F_{_{\text{W}}}^{\theta  \varphi }=-\tfrac{B \cot\theta }{r^2}+\tfrac{2 Q a (f_2+M r) \cot  \theta }{M r^6}.\nn
\end{align}
Now if we set $f_1=M$ in~\eqref{rc6}, we see that the terms proportional to $M$ in both expressions~\eqref{rc6} and~\eqref{rc7} are the same. In the case of the Reissner-Nordstr\"om black hole with $a=0$ and $f_2=-Q^2/2$, however, Eq.~\eqref{rc7} contains an extra term, $-Q^3/(Mr^3)$, in $F^{t r}$ and in the case of the Kerr-Newman black hole with $a\neq 0$ and $f_2=-Q^2/2$, Eq.~\eqref{rc7} fails to produce the term, $BQ^2\cot\theta/r^4$, in  $F^{\ta\vp}$. Even if $B$ is taken as a test field, this last term cannot be neglected near the axis of symmetry and in the vicinity of the horizon. As shown in~\eqref{force}, this produces the non-negligible extra magnetic force on a charged particle
\begin{equation}
-\tfrac{qQ^2B\cot\ta}{r^2}~\tfrac{\dd \ta}{\dd s}.
\end{equation}
This shows that the case of charged black holes lies beyond the realm of applicability of Wald's formula~\eqref{csc}.

In the case of the Kerr-Newman black hole, besides what we just mentioned in the previous paragraph, Wald's formula~\ref{csc} produces wrong and extra terms proportional to $f_2a=-Q^2a/2$ in $F^{t r}$, $F^{r\vp}$, and $F^{\ta\vp}$, and it yields a vanishing $F^{t\ta}$. Thus, the formulas~\eqref{rc6} modify greatly the expressions of the forces acting on a charged particle and add extra terms to them. For instance, the force $q F_{\si}{}^{\vp}(\dd x^{\si}/\dd s$ will still have the third extra term in~\eqref{force} and the force $q F_{\si}{}^{\ta}(\dd x^{\si}/\dd s$ will have two extra terms due to $F^{t\ta}$:
\begin{equation}
qF^{t\ta} \big(g_{tt}\tfrac{\dd t}{\dd s}+g_{t\vp}\tfrac{\dd \vp}{\dd s}\big).
\end{equation}
Dropping the second term proportional to $a^2$, there remains the term
\begin{equation}\label{rc8}
    -\tfrac{qQ^2B a  \sin  2 \theta }{2r^4}g_{tt}\tfrac{\dd t}{\dd s},
\end{equation}
which, for all $\ta$ and $g_{tt}>0$, drives positively (resp. negatively) charged particles to (resp. away from) the axis of symmetry.

Other more important applications will be given elsewhere~\cite{CME}.

\subsection*{Electromagnetic fields around rotating black holes}
As mentioned in Sec.~\ref{sec1}, the electromagnetic field is taken as a test field in a given background metric; that is, the presence of the field does not modify the background metric. This approximation results in the two constraints~\eqref{b1} and~\eqref{b2}.

To the best of our knowledge, there are no available exact, asymptotically flat, solutions to electromagnetic fields around (non)rotating black holes where the stress-energy of the electromagnetic field is taken into consideration except the well-known Kerr and Kerr-Newman solutions. Among existing approximate solutions, we find the stationary axisymmetric given in Refs.~\cite{JP,HR,CW,BL,IS}. Exact solutions with a Melvin asymptotic behavior do, however, exist~\cite{Ernst1,Ernst2}.

In Ref.~\cite{JP}, stationary axisymmetric electromagnetic fields surrounding a Kerr black hole were determined analytically. It was implicitly assumed that the stress-energy of the electromagnetic field is negligible not to affect the geometry of the background Kerr metric. Recall that the Kerr black hole has $f(r)=Mr$ ($f_1=M$ and $f_2=0$), so these field solutions do not extend to include the Kerr-Newman black hole nor more general black holes of the form~\eqref{m1} with $f_2\neq 0$. In order to compare these field solutions with~\eqref{fld1} we take $Q=0$ in Eq.~(29) of Ref.~\cite{JP}, which is the charge of the electromagnetic source (the background Kerr black hole is uncharged) and take $Q=0$ in~\eqref{fld1}, which is the charge of the background black hole. Now, if we fix the remaining constants in Eq.~(29) of Ref.~\cite{JP} such that $\al_1^i=-B\sqrt{M^2-a^2}/2$ ($\simeq -BM/2$ if $a$ small), $\al_l^r=0$, $\al_l^i=0$ ($l\neq 1$), $\bt_l^r=0$, and $\bt_l^i=0$, this produces the leading terms proportional to $r$ and $r^2$ in~\eqref{fld1}.

\section{Conclusion \label{secc}}

We have derived expressions for the vector potential and electromagnetic field of a rotating and nonrotating charged black hole immersed in a uniform magnetic field. The expressions are exact within the linear approximation and include all linear terms in both the rotation parameter and the magnetic field, thus introducing corrections to Wald's formulas. Since we have considered exact background black hole solutions, no special assumptions constraining the electric charge and the mass have been made. We have, however, made the implicit assumptions that the mass, electric charge, and rotation parameter are such that the solution is a black hole (not a naked solution without horizon(s)). The expressions apply to a variety of vacuum and nonvacuum solutions provided they satisfy asymptotically the linear Maxwell field equations.

As a first application we have observed the emergence of new extra force terms and evaluated some of them. Other applications will follow~\cite{CME}. The generalization along with the corrections made to Wald's formulas are crucial for a consistent analysis of (un)charged-particle dynamics around black holes~\cite{B2}-\cite{B4}, \cite{M1}-\cite{M4}, to mention but a few. From this point of view, some particle-dynamics analyses made in the literature have relied on Wald's formulas in cases where they do not apply.

The Kerr-Sen and Kerr-Newman-Taub-NUT black holes are charged solutions having dilaton fields resulting from dimensional reductions. Due to dilatons, the vacuum electromagnetic field equations~\eqref{VME} no longer describe the motion of $F_{\mu\nu}$. Moreover, their metrics diverge from the form~\eqref{m1}. It is another interesting topic to determine the vector potential of these black holes when they are immersed in a uniform magnetic field parallel to the axis of symmetry.

\section*{Acknowledgment} I thank B.~Ahmedov and M.~Jamil for suggesting a list of references.

%



\end{document}